\documentclass[
showpacs,preprintnumbers,
 amsmath,amssymb,
 aps,
 prd,
longbibliography,
floatfix,
lengthcheck, nofootinbib,
]{revtex4-1}

\usepackage[T1]{fontenc} 
\usepackage{lmodern}
\usepackage{graphicx}
\usepackage{bm}
\usepackage{amsmath}
\usepackage{amssymb}
\usepackage{color}

\begin{document}

\title{GR Effects in Supernova Neutrino Flavor Transformation}
\author{Yue Yang}
\author{James P.~Kneller}
\affiliation{Department of Physics, North Carolina State University, Raleigh, NC 27695 USA}

\date{\today}

\begin{abstract}
The strong gravitational field around a proto-neutron star can modify the neutrino flavor transformations that occur above the neutrinosphere via three General Relativistic (GR) effects: time dilation, energy redshift, and trajectory bending. Depending on the compactness of the central object, the neutrino self-interaction potential is up to three times as large as that without GR principally due to trajectory bending which increases the intersection angles between different neutrino trajectories, and time dilation which changes the fluxes. We determine whether GR effects are important for flavor transformation during the different epochs of a supernova by using multi-angle flavor transformation calculations and consider a density profile and neutrino spectra representative of both the accretion and cooling phases. We find the GR effects are smaller during the accretion phase due to low compactness of the proto-neutron star and merely delay the decoherence; the neutrino bipolar oscillations during the cooling phase are also delayed due to the GR effects but the delay may be more important because the delay occurs at radii where it might alter the nucleosynthesis in the neutrino driven wind.

\end{abstract}
\medskip

\maketitle


\section{Introduction}
\label{sec:intro}

The collapse of the core of a massive star at the end of its life forms a hot and dense object known as a proto-neutron star which cools via the emission of neutrinos over a period of $\sim 10\;{\rm s}$ \cite{2002RvMP...74.1015W,2007PhR...442...38J}. The spectra and flavor distribution of the neutrinos that emerge from the supernova are not the same as those emitted from the proto-neutron star: for a recent review see Mirizzi \emph{et al.} \cite{2016NCimR..39....1M}. At the present time the most sophisticated calculations of the neutrino flavor transformation adopt the so-called `bulb' model: the neutrino source is a spherically symmetric, hard neutrinosphere, the calculation assumes a steady state, and neutrinos are followed along multiple trajectories characterized by their angle of emission relative to the radial direction - the `multi-angle' approach \cite{Duan:2006an,Duan:2006jv}. The Hamiltonian governing the flavor evolution for a single neutrino depends on the local density profile plus a contribution from all the other neutrinos which are escaping the proto-neutron star - the neutrino self-interaction. The neutrino self-interaction depends upon the neutrino luminosity, mean energy and a term proportional to $1 - \cos\Theta$ due to the current-current nature of the weak interaction where $\Theta$ is the angle between two neutrino trajectories. Curiously, while the density profile and the neutrino spectra are sometimes taken from hydrodynamical simulations of supernova which include General Relativistic (GR) effects either exactly or approximately, e.g. from the simulations by Fischer \emph{et al.} \cite{2010A&A...517A..80F}, the calculations of the neutrino flavor transformation ignore them. 

The flavor transformation that occurs in a supernova will alter the expected signal from the next Galactic supernova \cite{2008PhRvD..77d5023K,2009PhRvL.103g1101G,2011JCAP...10..019V,2013PhRvD..88b3008L}, as well as modify the Diffuse Supernova Neutrino Background  \cite{2008JCAP...09..013C,2010ARNPS..60..439B,2010PhRvD..81e3002G,2011PhLB..702..209C,2012JCAP...07..012L,2013PhRvD..88h3012N,2016APh....79...49L}, and the nucleosynthesis that occurs in the neutrino driven wind \cite{1994ApJ...433..229W,2010JCAP...06..007C,2011JPhG...38c5201D,2015ApJ...808..188P,2015PhRvD..91f5016W}. Neutrino heating in the region behind the shock is thought to be the mechanism by which the star explodes and such heating depends upon the neutrino spectra of each flavor which depends upon the flavor transformation \cite{1985ApJ...295...14B,1996A&A...306..167J}. With so many different consequences of flavor transformation, one wonders how including GR in the flavor transformation calculations might alter our expectations. 

GR effects upon neutrino oscillations in vacuum have been considered on several occasions e.g.\ \cite{1979GReGr..11..391S,1996GReGr..28.1161A,1996PhRvD..54.1587P,1997PhRvD..56.1895F,1998PThPh.100.1145K,1999PhRvD..59f7301B,2005PhRvD..71g3011L,2015GReGr..47...62V}. The inclusion of matter is occasionally considered \cite{1997PhRvD..55.7960C,2004GReGr..36.2183A,2013JCAP...06..015D,2016NuPhB.911..563Z} and the effect of GR usually limited to a shift in location and adiabaticity of the Mikheyev-Smirnov-Wolfenstein (MSW) resonance \cite{Mikheyev:1985aa,1978PhRvD..17.2369W} via the redshift of the neutrino energy. The effects of GR upon neutrino self-interactions have not been considered. The effect of GR has also been studied for the neutrinos emitted from the accretion disk surrounding a black hole formed in the merger of two neutron stars, a black hole and a neutron star, or in a collapsar. For example, Caballero, McLaughlin and Surman \cite{2012ApJ...745..170C} studied the GR effects for accretion disk neutrinos (but without neutrino transformation) and found the effects upon the nucleosynthesis were large because of the significant changes to the neutrino flux. 

The aim of this paper is to explore the GR effects upon flavor transformation in supernovae including neutrino self-interactions and determine whether they might be important in different phases of the explosion. Our paper is organized as follows. In section \S\ref{sec:description} we describe our calculation and how the GR effects are included. Section \S\ref{sec:results} contains our results for the two representative cases we study: luminosities, mean and rms energies, density profiles and source compactness characteristic of the accretion phase, and a different set representative of the cooling phase. In section \S\ref{sec:halo} we discuss the conditions that lead to the formation of a neutrino halo - neutrinos that were emitted but which later turned-around and returned to the proto-neutron star. We present a summary and our conclusions in section \S\ref{sec:conclusions}.


\section{Calculation Description}
\label{sec:description}

\subsection{GR Effects Upon Neutrinos}

Before describing the formulation of neutrino oscillations in a curved spacetime, we first describe the three general relativistic effects that will be important. For this paper we adopt an exterior Schwarzschild metric for the space beyond the neutrinosphere\footnote{For simplicity we ignore the gravitational effect of the matter outside the neutrinosphere.} which is given by
\begin{equation}\label{eqn:metric1}
d{\tau^2} = B\left(r\right)dt^{2} - \frac{dr^{2}}{B\left(r\right)} - r^{2}d\psi^{2} - r^{2}{\sin^2}\psi\,d\phi^{2},
\end{equation}
where the function $B(r)$ is $B\left(r\right) = 1 - r_s/r$ and $r_s$ is the Schwarzschild radius given by $r_s = 2 G M$ with $M$ the gravitational mass. Throughout our paper we set $\hbar=c=1$. Since the rest mass of all neutrino species are much smaller than the typical energies of supernova neutrinos, we can comfortably take the ultra-relativistic limit and assume neutrinos follow null geodesics just like photons. The Schwarzschild metric is isotropic so all geodesics are planar. By setting $d{\tau^2} = 0$ and $d\phi=0$ so that the geodesic lies in the plane perpendicular to the equatorial plane, we obtain 
\begin{equation}\label{eqn:metric2}
B\left(r\right)dt^{2} = \frac{dr^{2}}{B\left(r\right)} + r^{2}d{\psi^2}.
\end{equation}
The energy of a neutrino $E$ decreases as it climbs out of the gravitational well such that its energy at a given radial coordinate $r$ relative to its energy at $r\rightarrow \infty$, $E_{\infty}$, is 
\begin{equation}
\frac{E}{E_{\infty}}= \frac{1}{\sqrt{B\left(r\right)}}.
\end{equation}
The angular momentum $\ell$ of the neutrino also decreases as it climbs out of the potential well by the same scaling. This means the ratio of the neutrino's angular momentum to its energy is constant and in our chosen plane is given by 
\begin{equation}\label{eqn:metric3}
\frac{\ell}{E} = \frac{r^2}{B\left(r\right)} \left|\frac{d\psi}{dt}\right| = b 
\end{equation}
where $b$ is a constant called the \emph{impact parameter}. The impact parameter can be evaluated at the neutrinosphere $r=R_{\nu}$ where we find it is given by 
\begin{equation}
b = \frac{R_{\nu}\sin\theta_R}{\sqrt{1 - r_s/R_{\nu}} } ,
\end{equation}
where $\theta_R$ is the the emission angle of the neutrino with respect to the radial direction at the neutrinosphere. Using Eq. (\ref{eqn:metric3}) to eliminate $dt$ from Eq. (\ref{eqn:metric2}) we find\footnote{Here the plus sign is for outgoing neutrinos, the minus sign is for ingoing neutrinos, this is true for all following equations.}  
\begin{equation}
d\psi  = \pm{\left[ {\frac{1}{b^2} - \frac{1}{r^2}\,B\left( r \right)} \right]^{-1/2}}\frac{dr}{r^2},
\end{equation}
This equation can be used to describe the neutrino trajectory associated with a certain emission angle $\theta_R$. Or using Eq. (\ref{eqn:metric3}) to eliminate $d\psi$ from Eq. (\ref{eqn:metric2}) gives
\begin{equation} \label{eqn:dt}
dt = \pm\frac{1}{B\left(r\right)}\frac{dr}{\sqrt {1 - \frac{b^2}{r^2}B\left(r\right)} }.
\end{equation}
For an observer at position $r$ the relation between the coordinate time $t$ and the local proper time\footnote{The ``local proper time'' is defined as the clock time of an observer sitting at a particular point along the neutrino trajectory.} $\tau$ is 
\begin{equation} 
d\tau^2 = B\left(r\right) dt^2
\end{equation} 
so using the result from Eq. (\ref{eqn:dt}) we find 
\begin{equation}\label{eqn:proper time}
d\tau = \pm\frac{1}{\sqrt{B\left(r\right)}}\frac{dr}{\sqrt{1 - \frac{b^2}{r^2}B\left(r\right)} }.
\end{equation}
This collection of equations will be useful when we describe flavor oscillations in a curved spacetime.

\subsection{Neutrino Oscillations In A Curved Spacetime}

\begin{figure}[b]
\centering
\includegraphics[scale=0.6]{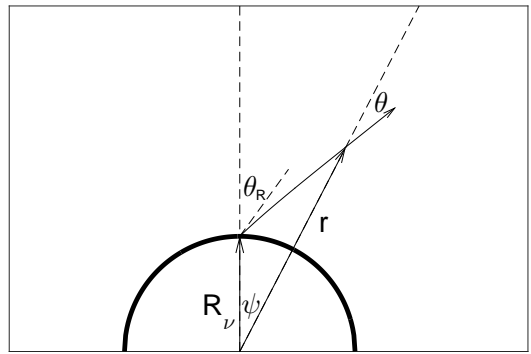}
\caption{The schematic of a neutrino trajectory in strong gravitational field. Here $R_{\nu}$ is the radius of neutrinosphere, $r$ is the distance from the center, $\theta_R$ is the emission angle, $\psi$ is the polar angle, and $\theta$ is the angle of intersection.}
\label{fig:angles}
\end{figure}
Our calculations of the effects of GR on neutrino flavor transformation are based upon the neutrino bulb model established by Duan \emph{et al.} \cite{Duan:2006an,Duan:2006jv}. In this model, neutrinos are emitted from a hard neutrinosphere with radius $R_{\nu}$ and for simplicity we assume the angular distribution of emission is half-isotropic. The setup is illustrated in Fig. \ref{fig:angles} which shows the trajectory of a neutrino emitted at the neutrinosphere $R_{\nu}$ with angle $\theta_{R}$ relative to the radial direction. After propagating to radial coordinate $r$ with angle $\psi$ relative to the radial direction at the point of emission, it makes an angle $\theta$ relative to the radial direction at $(r, \psi)$. The formulation of neutrino flavor transformation in a curved spacetime has been considered on multiple occasions \cite{1997PhRvD..56.1895F,1998PThPh.100.1145K,1999PhRvD..59f7301B,2005PhRvD..71g3011L,2015GReGr..47...62V,1997PhRvD..55.7960C}. The flavor state at some local proper time $\tau$ of a neutrino with momentum ${\bf q}$ is related to the flavor state at the local proper time of emission $\tau_0$ with momentum ${\bf q}_0$ via an evolution matrix $S(\tau,{\bf{q}};\tau_0,{\bf q}_0)$ which evolves according to the Schr\"odinger equation. In a curved spacetime the evolution matrices evolves with the local proper time $\tau$ as 
\begin{equation}
i\frac{dS}{d\tau} = H\left(\tau\right)S.
\end{equation}
Here $H$ is the Hamiltonian which is also a function of the local proper time for the case of neutrinos in a non-uniform medium. The local proper time $\tau$ may be replaced with the radial coordinate $r$ by using Eq. (\ref{eqn:proper time}) once the impact parameter/emission angle is given. Similarly, the evolution of the antineutrinos is given by an evolution matrix $\bar{S}$ which evolves according to a Hamiltonian $\bar{H}$. Once the evolution matrix has been found, the probability that a neutrino in some generic initial state $\nu_{j}$ with momentum ${\bf q_0}$ at $\tau_0$ is later detected as state $\nu_i$ at proper time $\tau$ and momentum ${\bf q}$ is $P(\nu_j \rightarrow \nu_i) = P_{ij} = |S_{ij}(\tau,{\bf{q}};\tau_0,{\bf q}_0)|^2$. 

The Hamiltonian $H$ is the sum of three terms: $H = H_V + H_{M} + H_{SI}$, where $H_{V}$ is the vacuum term, $H_M$ is the matter term to describe the effect of passing through matter, and $H_{SI}$ is a term due to neutrino self-interactions. For the antineutrinos the Hamiltonian is also a sum of three terms with $\bar{H} = \bar{H}_V + \bar{H}_{M} + \bar{H}_{SI}$, which are related to the corresponding terms in the neutrino Hamiltonian via $\bar{H}_V = H_{V}^{\ast}$, $\bar{H}_M = -\bar{H}^{\ast}_M$, $\bar{H}_{SI} = -\bar{H}^{\ast}_{SI}$. In a flat spacetime the vacuum term for a neutrino with energy $E$ takes the form of
\begin{equation}
H^{(f)}_{V} = \frac{1}{2E}\,U_V \left( \begin{array}{*{20}{c}} m_1^2 & 0 & 0  \\ 0 & m_2^2 & 0  \\ 0 & 0 & m_3^2 \end{array} \right) U_V^{\dagger}
\end{equation}
\\
where $m_i$ are the neutrino masses and $U_V$ is the unitary matrix relating the `mass' and flavor bases. The flavor basis is denoted by the superscript $(f)$ upon relevant quantities and we order the rows/columns as $e$, $\mu$, $\tau$ (here $\tau$ is the neutrino flavor, not local proper time). We adopt the Particle Data Group parameterization of the matrix $U_V$ which is in terms of three mixing angles $\theta_{12}$, $\theta_{13}$ and $\theta_{23}$ plus a CP violating phase $\delta_{CP}$ \cite{Olive:2016xmw}. In a curved spacetime the energy of a neutrino is dependent on position due to the gravitational redshift so the vacuum term will change accordingly and is
\begin{equation}
H^{(f)}_{V} = \frac{\sqrt{B(r)}}{2E_{\infty}}\,U_V \left( \begin{array}{*{20}{c}} m_1^2 & 0 & 0  \\ 0 & m_2^2 & 0  \\ 0 & 0 & m_3^2 \end{array} \right) U_V^{\dagger}.
\end{equation}
\\
The matter Hamiltonian $H_M$ in the flavor basis depends upon the electron density $n_{e}(r)$ and is simply 
\begin{equation}
H^{(f)}_{M} = \sqrt{2}\,G_F\,n_e(r) \left( \begin{array}{*{20}{c}} 1 & 0 & 0  \\ 0 & 0 & 0  \\ 0 & 0 & 0 \end{array} \right).
\end{equation}

\subsection{The GR correction to neutrino self-interactions}

In addition to the vacuum and matter terms, in a neutrino dense environment such as a supernova we must add to the Hamiltonian a term due to neutrino self-interactions. The form of the self-interaction is
\begin{widetext}
\begin{equation}
\label{eqn:self-coupling}
{ H_{SI}}\left( {r,{\bf{q}}} \right) = \sqrt 2 {G_F}\sum\limits_{\alpha  = e,\mu ,\tau } \int \left( 1 - {\bf{\hat q}} \cdot {\bf{\hat q}}' \right)\left[ \rho_{\alpha }\left( r,{\bf q}' \right)\,dn_{\alpha}\left( r,{\bf q}' \right)  - \rho _{\bar\alpha}^*\left( r,{\bf q}' \right)\,dn_{\bar\alpha}\left( r,{\bf q}' \right) \right]\,dq'
\end{equation}
\end{widetext}
where $\rho_{\alpha}(r,{\bf q})$ is the density matrix of the neutrinos at position $r$ with momentum ${\bf q}$ and initial flavor $\alpha$ defined as $\rho_{\alpha}(r,{\bf q})=\psi_{\alpha}(r,{\bf q})\psi^{\dag}_{\alpha}(r,{\bf q})$, with $\psi_{\alpha}(r,{\bf q})$ being the corresponding normalized neutrino wave function, $dn_{\alpha}(r,{\bf q})$ is the differential neutrino number density \cite{Duan:2006an}, which is the differential contribution to the neutrino number density at $r$ from those neutrinos with initial flavor $\alpha$ and energy $\left|{\bf q}\right|$ propagating in the directions between ${\bf{\hat q}}$ and ${\bf {\hat q}}+d{\bf{\hat q}}$, per unit energy (the hats on ${\bf q}$ and ${\bf q}'$ indicate unit vectors). Note that here we have replaced the local proper time $\tau$ with the radial coordinate $r$ to denote the location along a given neutrino trajectory.  

\begin{figure}[t]
\includegraphics[scale=0.5]{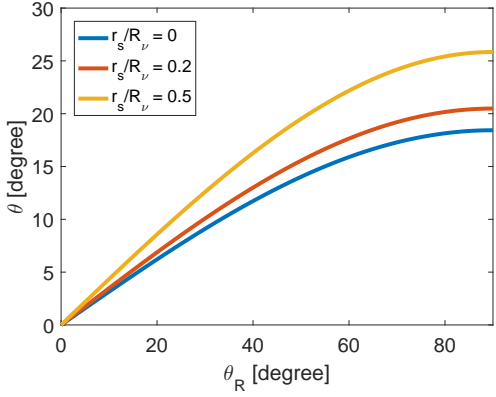}
\caption{The relationship between $\theta$ and $\theta_R$ for $r_s/R_{\nu}=0,0.2\;\rm{and}\; 0.5$ evaluated at $r=10\,R_{\nu}$.}
\label{fig:theta_vs_thetaR}
\end{figure}

In order to use Eq. (\ref{eqn:self-coupling}) we have to first specify the expression for $dn_{\alpha}(r,{\bf q})$. This requires relating the neutrino momenta ${\bf q}$ at radial coordinate $r$ back to their values ${\bf q}_0$ at the neutrinosphere where they are initialized. After this relationship is obtained we can substitute $dn_{\alpha}({r,\bf q})$ with $dn_{\alpha}(R_{\nu},{\bf q_0})$ and calculate $H_{SI}$ by integrating over the neutrino momentum distributions at the neutrinosphere. While the magnitude of ${\bf q}$ is related to the magnitude of ${\bf q_0}$ via an energy redshift $q=q_0\sqrt{B(R_{\nu})/B(r)}$, relating $\hat{\bf q}$ to $\hat{\bf q}_0$ means finding the relation between the emission angle $\theta_R$ and the angle $\theta$ shown in Fig. \ref{fig:angles} since the neutrino trajectory is planar.  In flat spacetime, the relation between $\theta_R$ and $\theta$ can be found through geometric arguments \cite{Duan:2006an}. In a curved spacetime, however, $\theta$ and $\theta_R$ might be expected to be related only after solving for the neutrino trajectory. But fortunately, for the Schwarzschild metric the relation between $\theta$ and $\theta_R$ can also be found simply by making use of the fact that the impact parameter $b$ is a conserved quantity along each neutrino trajectory \cite{2012ApJ...745..170C}. It makes no difference whether the impact parameter is evaluated at $R_{\nu}$ or at $r$, therefore $b(r)=b(R_{\nu})$. Using this conserved quantity we must have 
\begin{equation}\label{eqn:b_conservation}
\frac{r\sin \theta}{\sqrt{1 - r_s/r}} = \frac{R_{\nu}\sin\theta_R}{\sqrt{1 - r_s/R_{\nu}} },
\end{equation}
from which we find
\begin{equation}\label{eqn:cos(theta)}
\cos\theta = \sqrt{1 - \left( \frac{R_{\nu}\sin\theta_R}{r}  \right)^2 \left( \frac{1 - r_s/r}{1 - r_s/R_{\nu} } \right)} .
\end{equation}
In Fig. \ref{fig:theta_vs_thetaR} we plot the angle $\theta$ as a function of emission angle $\theta_R$ for three different ratios of $r_s$ to $R_{\nu}$ at $r=10\,R_{\nu}$. The figure shows that for each particular emission angle $\theta_R$, the trajectory bending effect always makes the angle $\theta$ larger than without GR. In the bulb model $\left( 1 - {\bf{\hat q}} \cdot {\bf{\hat q}}' \right)$ is found to be equivalent to $\left( 1 - \cos\theta \cos\theta' \right)$ after averaging over the angles in the plane perpendicular to the radial direction. Thus the correction to $\cos\theta$ by GR increases the magnitude of $H_{SI}$ by increasing the value of $1 - {\bf{\hat q}} \cdot {\bf{\hat q}}'$ for every neutrino. 

Now we have the expression relating $\theta$ to $\theta_R$, we can write the expression for the differential number density, after taking time dilation into account, as 
\begin{widetext}
\begin{equation}
\label{eqn:dn_alpha}
dn_{\alpha}\left(r,{\bf q}\right)\equiv dn_{\alpha}\left(r,q,\theta\right)\equiv dn_{\alpha}\left(R_{\nu},q_0,\theta_R\right) = \frac{1}{2\pi r^{2}\sqrt{B(r)}}\,\left[ \frac{L_{\alpha,\infty}}{\left\langle E_{\alpha,\infty} \right\rangle} \right]\,f_{\alpha}\left( q_0 \right)\left(\frac{\cos\theta_R}{\cos\theta}\right)\left(\frac{dq_0}{dq}\right)\,d\cos\theta_{R},
\end{equation}
\end{widetext}
where $f_{\alpha}\left( q_0 \right)$ is the normalized distribution function for flavor $\alpha$ with momentum $q_0$ that redshifts to $q$ at $r$, $L_{\alpha,\infty}$ is the luminosity of flavor $\alpha$ at infinity if no flavor transformation had occurred, and similarly $\left\langle E_{\alpha,\infty}\right\rangle$ is the mean energy of neutrinos of flavor $\alpha$ at infinity again assuming no flavor transformation had occurred. The expression for the antineutrinos is similar. The derivation of Eq. (\ref{eqn:dn_alpha}) can be found in the Appendix.

\begin{figure}[t]
\centering
\includegraphics[scale=0.6]{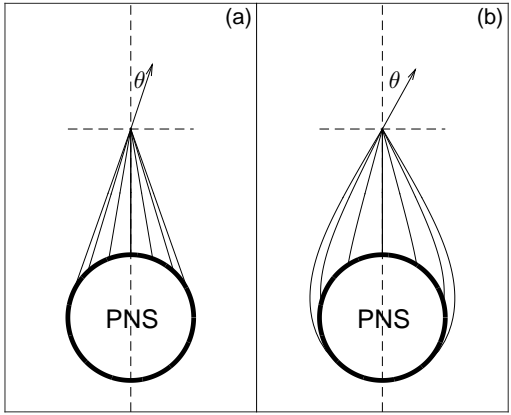}
\caption{The neutrino trajectories converging at $r=3R_{\nu}$ for (a) $r_s/R_{\nu}=0$ and (b) $r_s/R_{\nu}=0.6$.}
\label{fig:enlarged_angle}
\end{figure}

The density matrix $\rho_{\alpha}(r,{\bf q})$ for neutrinos at $r$ with momentum ${\bf q}$ is related to the corresponding density matrix at the neutrinosphere via $\rho_{\alpha}(r,{\bf q}) =S(r,{\bf q};R_{\nu},{\bf q}_0)\,\rho_{\alpha}(R_{\nu},{\bf{q}}_0)\,S^{\dag}(r,{\bf q};R_{\nu},{\bf q}_0)$ and the same for the antineutrinos using the evolution matrix $\bar{S}(r,{\bf q};R_{\nu},{\bf q}_0)$. 

Combining these equations together, we obtain the GR corrected expression of neutrino self-interaction in curved spacetime as
\begin{widetext}
\begin{eqnarray}\label{eqn:GR_self_coupling}
H_{SI}\left( {r,{\bf{q}}} \right) & = & \frac{\sqrt{2}\,G_{F}}{2\pi r^{2}\,\sqrt{B(r)}} \sum\limits_{\alpha  = e,\mu ,\tau } \nonumber \\ 
& & \times \int \left( 1 - \cos\theta \cos\theta' \right) \left\{ \left[\frac{L_{\alpha,\infty}}{\left\langle E_{\alpha,\infty}\right\rangle} \right]\, \,\rho_{\alpha}(r,{\bf q'}) \, f_{\alpha}\left(q'_0 \right)  - \left[\frac{L_{\bar\alpha,\infty}}{\left\langle E_{\bar\alpha,\infty}\right\rangle } \, \right]\, \rho^{\star}_{\bar\alpha}(r,{\bf q'})  \,f_{\bar\alpha}\left(q'_0 \right) \right\} \left(\frac{ \cos\theta_{R}'}{ \cos\theta' }\right)\,d\cos\theta_{R}'\,dq'_0. \nonumber \\ & & 
\end{eqnarray}
\end{widetext}

When we take the weak gravity limit $r_s \ll r$ and $r_s \ll R_{\nu}$ we find this expression reduces to the same equation found in Duan \emph{et al.} \cite{Duan:2006an,Duan:2006jv}. This equation includes two GR effects: trajectory bending and time dilation (the energy redshift of the luminosity cancels with the energy redshift of the mean energy). In order to appreciate how significant the GR effects can be for the self-interaction Hamiltonian we show in Fig. \ref{fig:enlarged_angle} the neutrino trajectories which converge at a certain point above the surface of the central proto-neutron star. From the perspective of an observer at this point, the neutrinos seem to be coming from an expanded source whose radius is increased by a factor of $\sqrt{(1-r_s/r)/(1-r_s/R_{\nu})}$, which can be seen from Eq. (\ref{eqn:cos(theta)}). As noted earlier, the effect of trajectory bending causes the neutrino trajectories to cross at larger angles than in the case without GR. Time dilation also enhances the self-interaction because it leads to a larger effective neutrino flow rate. Close to the neutrinosphere time dilation is the larger effect because the effect of trajectory bending is small. At larger radii the situation is reversed with trajectory bending more important than time dilation. 

\begin{figure}
\includegraphics[scale=0.5]{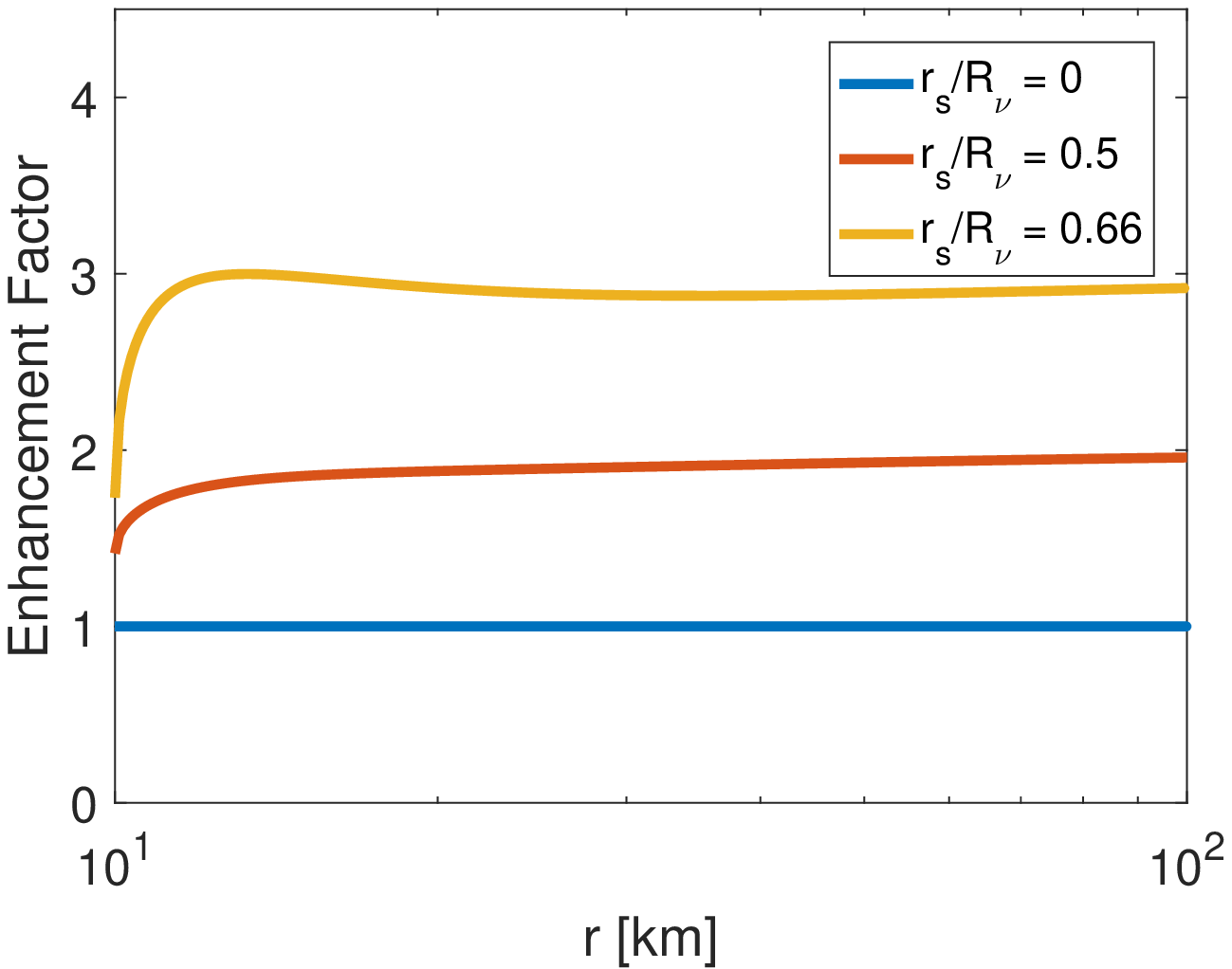}
\centering
\includegraphics[scale=0.5]{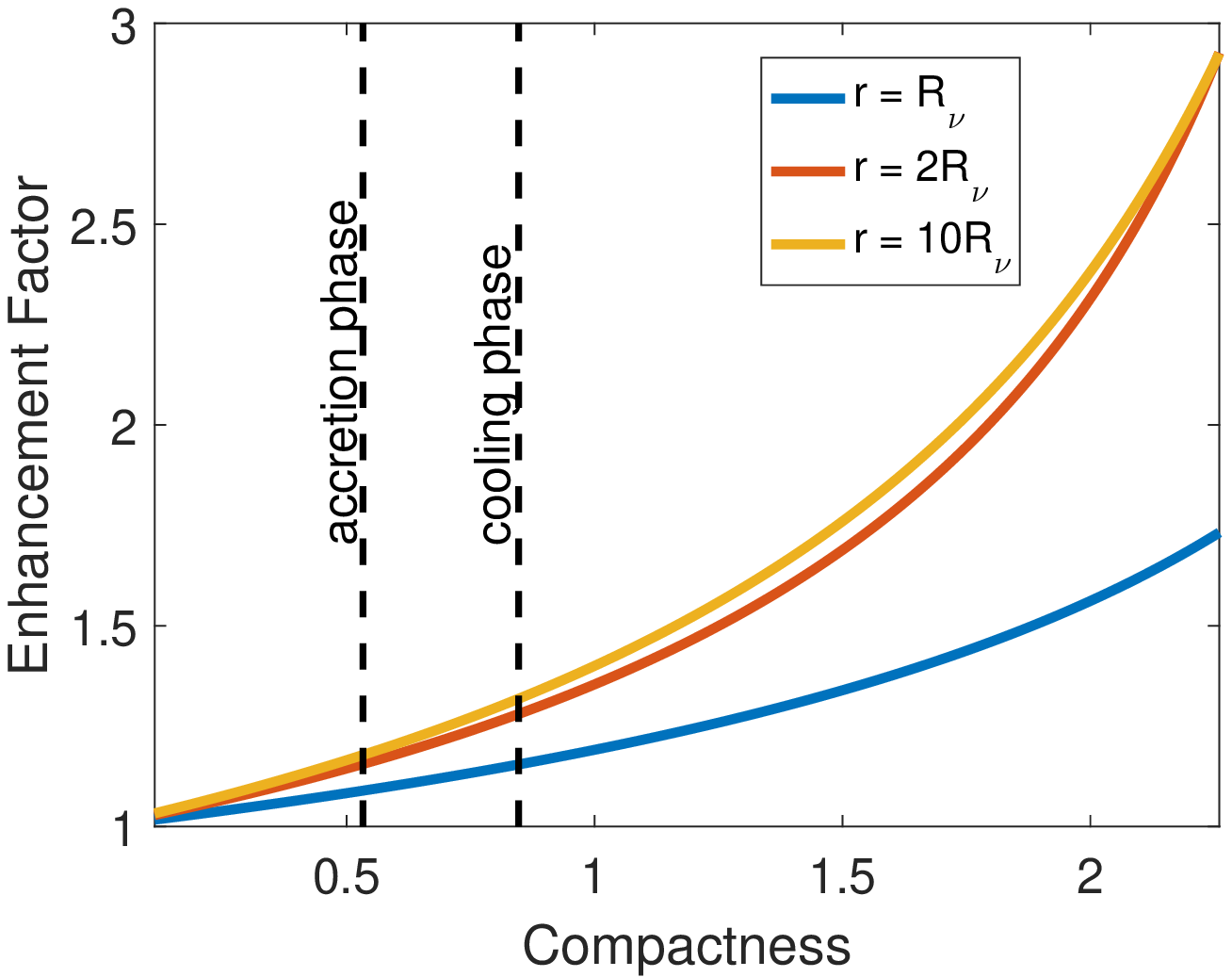}
\caption{\emph{Top:}
The enhancement factor as a function of distance for three different ratios of the Schwarzschild radius relative to the neutrinosphere radius. \emph{Bottom:} The enhancement factor as a function of compactness, at three different distances. The two vertical dashed lines indicate the compactness of the sources we use in our calculations for the accretion phase and cooling phase, respectively. \label{fig:enhancement_factor} }
\end{figure}

To quantify the magnitude of the GR effects upon the self-interaction we show in the top panel of Fig. \ref{fig:enhancement_factor} the enhancement of the self-interaction due to GR, which is defined to be the ratio of the magnitude of the self-interaction potential with GR effects to that without, as a function of the coordinate $r$ and assuming no flavor oscillation occurs, for different values of $r_s/R_{\nu}$. The striking feature of the GR effects is that, even though the spacetime curvature is only pronounced near the proto-neutron star, the enhancement of the neutrino self-coupling turns out to be a long-range effect that is asymptotic to a value greater than unity which depends upon the ratio $r_s/R_{\nu}$. Since the influence of GR on neutrino flavor transformation is not just a local effect, it can have repercussions upon processes at larger radii such as neutrino heating in the accretion phase and nucleosynthesis in the cooling phase.

As we have seen, the magnitude of the GR effect is governed by ratio of the radius of the neutrinosphere relative to the Schwarzschild radius of the proto-neutron star which itself is just proportional to the mass of the proto-neutron star. This suggests we define a neutrino `compactness' - similar to the definition of compactness found in O'Connor \& Ott \cite{2013ApJ...762..126O} - as
\begin{equation}
\xi_{\nu} = \frac{M/M_{\odot}}{R_{\nu}/10{\;\rm{km}}} = \frac{r_s/2.95{\;\rm{km}}}{R_{\nu}/10{\;\rm{km}}}= 3.39\frac{r_s}{R_{\nu}}.
\end{equation}
In the bottom panel of Fig. \ref{fig:enhancement_factor} we plot the enhancement factor as a function of compactness at different distances from the center of the proto-neutron star. For a very compact neutrino source we find the enhancement of the self-interaction can be as large as a factor of $300\%$ if $\xi_{\nu} \sim 2.26$ which corresponds to $r_s/R_{\nu}=2/3$. We shall explain the significance of this compactness in section \S\ref{sec:halo}. The blue line in this figures shows the enhancement factor at the neutrinosphere, where the trajectory bending effect is minimal. Here the enhancement is purely due to time dilation.


\section{Numerical Calculations}
\label{sec:results}

With the formulation complete and with the insights gained from the computation of the enhancement as a function of compactness, we proceed to compute numerically the multi-angle neutrino flavor evolution for two representative cases. These are a density profile, neutrino spectra and compactness typical of the accretion phase of a supernova, and one representative of the cooling phase. The neutrino mixing angles and square mass differences we adopt are $m^2_2-m^2_1=7.5\times10^{-5}\;\text{eV}^2$, $m^2_{3}-m^2_{2}=-2.32\times10^{-3}\;\text{eV}^2$ $\theta_{12}=33.9^\circ$ $\theta_{13}=9^\circ$ and $\theta_{23}=45^\circ$. The CP phase $\delta_{CP}$ is set to zero. We do not consider a normal mass ordering on the basis of the results by Chakraborty \emph{et al.} \cite{2011PhRvL.107o1101C} and Wu \emph{et al.}  \cite{2015PhRvD..91f5016W}.

\subsection{Application to SN accretion phase}

\begin{figure}[t]
\includegraphics[clip,angle=0,width=\linewidth]{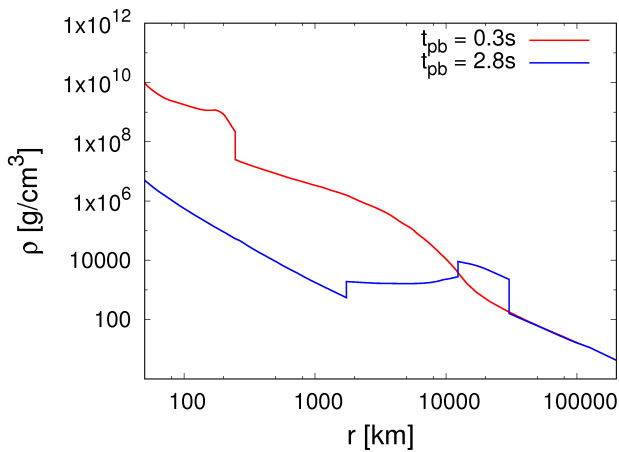}
\caption{The matter density profiles of the $10.8\;\rm{M_{\odot}}$ simulation by Fischer \emph{et al.} \cite{2010A&A...517A..80F} at postbounce times $t_{pb} = 0.3\;{\rm s}$ (red solid line) and $t_{pb} = 2.8\;{\rm s}$ (blue solid line).}
\label{fig:density}
\end{figure}

\begin{table}[b]
\begin{tabular}{lc}
Flavor\; & \;Luminosity $L_{\alpha,\infty}$ \; \\
\hline
$e$           & $41.52\times 10^{51}\;{\rm erg/s}$  \\  
$\mu$, $\tau$ & $14.23\times 10^{51}\;{\rm erg/s}$  \\  
$\bar{e}$     & $42.35\times 10^{51}\;{\rm erg/s}$  \\   
$\bar{\mu}$, $\bar{\tau}$ & $14.39\times 10^{51}\;{\rm erg/s}$  \\  
\\Flavor\; & \;Mean Energy $\langle E_{\alpha,\infty}\rangle$ \; \\
\hline
$e$           & $10.39\;{\rm MeV}$ \\  
$\mu$, $\tau$ & $16.19\;{\rm MeV}$ \\  
$\bar{e}$     & $12.67\;{\rm MeV}$ \\   
$\bar{\mu}$, $\bar{\tau}$ & $16.40\;{\rm MeV}$ 
\end{tabular}
\caption{The luminosities and mean energies used for the accretion phase calculation.} \label{tab:accretionphase}
\end{table}

For the accretion phase we use the density profile at $t_{pb}=0.3\;{\rm s}$ postbounce from Fischer \emph{et al.} \cite{2010A&A...517A..80F} for the $10.8\;{\rm M_{\odot}}$ progenitor. As previously stated, this simulation includes GR effects in both the hydrodynamics and evolution of the neutrino phase space density (see Liebend{\"o}rfer \emph{et al.} \cite{2004ApJS..150..263L} for further details about the code). The density profile at this snapshot time is shown by the red line in Fig. \ref{fig:density}. We set the neutrinosphere radius to be $R_{\nu}=25\;{\rm km}$ which corresponds to the minimum of the electron fraction for this model at this time. This working definition for the neutrinosphere radius comes from noting the coincidence of the electron fraction minimum and the neutrinosphere radii shown in figures (7) and (8) in Fischer \emph{et al.} and produces a curve which is similar to figure (15) found in their paper. We note that the value of $R_{\nu}$ we adopt is different from the value estimated by others, e.g.\ \cite{2012PhRvD..85k3007S,2011PhRvL.107o1101C}, which tend to use relatively larger values for $R_{\nu}$ during the accretion phase. From the simulation we find the mass enclosed within the $R_{\nu}=25\;{\rm km}$ radius is $M = 1.33\;{\rm M_{\odot}}$, giving a compactness of $\xi_{\nu} = 0.53$. The neutrino luminosities and mean energies we use are also taken from the same simulation and are listed in table (\ref{tab:accretionphase}). To save computational resources we use a source distribution $f_{\alpha}(q_0)$ which is a delta-function at a single energy taken to be $15\;{\rm MeV}$. Single-energy calculations were also undertaken by Chakraborty \emph{et al.} \cite{2011PhRvD..84b5002C} when they also studied the self-interaction effects during the accretion phase. As previously stated, the angular distribution is assumed to be half-isotropic which is the same distribution used in Duan \emph{et al.} \cite{Duan:2006an,Duan:2006jv}. 

\begin{figure}[t]
\includegraphics[clip,angle=270,width=\linewidth]{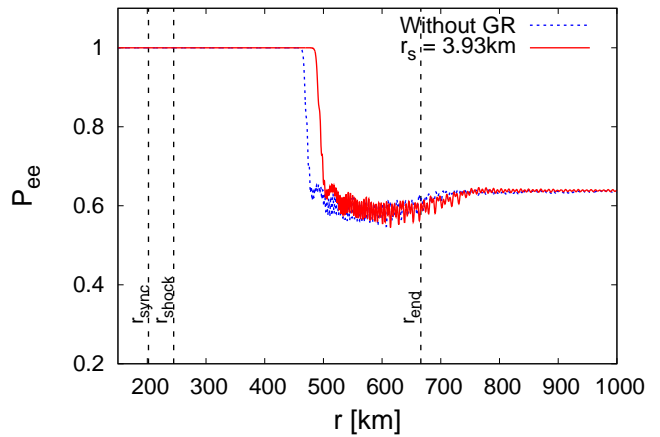}
\caption{The survival probability of electron neutrinos as a function of distance in the SN accretion phase, when $t_{pb}=0.3$ {\rm s}. The result is averaged over all angular bins. $R_{\nu}$ is set to $25$ {\rm km}, red solid line and blue dotted line are the results with and without GR effect, respectively. The vertical dashed lines labeled $r_{sync}$ and $r_{end}$ are the predicted beginning and ending locations of bipolar oscillations as given by the equations given in \cite{2011PhRvD..84b5002C}. The position of the shock wave is also indicated and labeled as $r_{shock}$.}
\label{fig:accretion_phase}
\end{figure}

Our results are shown in Fig. \ref{fig:accretion_phase} which is a plot of the electron flavor survival probability averaged over all angular bins as a function of distance. In the figure we also include three vertical dashed lines to indicate the start of the bipolar oscillation region, the position of the shockwave, and the end of the bipolar oscillation region. The predictions for the beginning and end of the bipolar oscillation region come from equations given in Chakraborty \emph{et al.} \cite{2011PhRvD..84b5002C}. The change in the angle-averaged survival probability $P_{ee}$ which occurs at $r\sim 475\;{\rm km}$ is simply decoherence \cite{2011PhRvL.107o1101C}. Comparing the results with and without GR effects we see the decoherence is slightly delayed when GR is included but the difference is only of order $\sim 20\;{\rm km}$ and the final result is identical to the case without GR. Thus it appears GR has little effect upon flavor transformation during the accretion phase and where little change occurs is in a region where it has little consequence. 

\subsection{Application to SN cooling phase}

As the proto-neutron star cools it contracts which increases the compactness. The sensitivity of the neutrino self-interaction to the compactness means we might expect a larger effect from GR during the cooling phase. To test whether this is the case we use the density profile at $t_{pb}=2.8\;{\rm s}$ postbounce from the Fischer \emph{et al.} \cite{2010A&A...517A..80F} simulation for the same $10.8\;{\rm M_{\odot}}$ progenitor and which is shown by the blue line in Fig. \ref{fig:density}. We set the neutrinosphere radius to be $R_{\nu}=17\;{\rm km}$ which, again, is close to the minimum of the electron fraction for this model at this time and consistent with figure (15) from Fischer \emph{et al.}. The mass enclosed within this radius is  $M\approx 1.44\;{\rm M_{\odot}}$, giving a compactness of $\xi_{\nu} = 0.85$. For this cooling epoch calculation we use multi-energy as well as multi-angle. The neutrino energy range is chosen to be $E_{\infty} = 1\;{\rm MeV}$ to $E_{\infty} = 60\;{\rm MeV}$, and is divided into 300, equally spaced, energy bins. To generate the neutrino spectra for flavor $\alpha$ at the neutrinosphere we use the luminosities, mean energies and rms energies at this snapshot of the simulation -  listed in table (\ref{tab:coolingphase}) - and insert them into the pinched thermal spectrum of Keil, Raffelt and Janka \cite{2003ApJ...590..971K} which has the form 
\begin{equation}
{f_\alpha }({q_0}) = \frac{{{{({A_\alpha } + 1)}^{{A_\alpha } + 1}}q_0^{{A_\alpha }}}}{{{{\langle {E_{\alpha ,{R_\nu }}}\rangle }^{{A_\alpha } + 1}}\Gamma ({A_\alpha } + 1)}}\exp \left( { - \frac{{({A_\alpha } + 1)\,{q_0}}}{{\langle {E_{\alpha ,{R_\nu }}}\rangle }}} \right),
\end{equation}
with $\langle E_{\alpha ,{R_\nu }}\rangle  = \langle E_{\alpha ,\infty}\rangle /\sqrt{B(R_{\nu})}$ and the pinch parameter $A_{\alpha}$ for flavor $\alpha$ is given by 
\begin{equation}
A_{\alpha} = \frac{2\,\langle E_{\alpha,\infty}\rangle^2 - \langle E^2_{\alpha,\infty}\rangle }{\langle E^2_{\alpha,\infty}\rangle - \langle E_{\alpha,\infty}\rangle^2 }.
\end{equation}

\begin{table}[b]
\begin{tabular}{lc}
Flavor\; & \;\;Luminosity $L_{\alpha,\infty}$  \\
\hline
$e$           & $2.504\times 10^{51}\;{\rm erg/s}$ \\  
$\mu$, $\tau$ & $2.864\times 10^{51}\;{\rm erg/s}$ \\  
$\bar{e}$     & $2.277\times 10^{51}\;{\rm erg/s}$ \\   
$\bar{\mu}$, $\bar{\tau}$ & $2.875\times 10^{51}\;{\rm erg/s}$ \\  
\\
Flavor\; & \;Mean Energy $\langle E_{\alpha,\infty}\rangle$\; \\
\hline
$e$           & $9.891\;{\rm MeV}$ \\  
$\mu$, $\tau$ & $12.66\;{\rm MeV}$ \\  
$\bar{e}$     & $11.83\;{\rm MeV}$ \\   
$\bar{\mu}$, $\bar{\tau}$ & $12.70\;{\rm MeV}$ \\  
\\
Flavor\; & \;rms Energy $\sqrt{ \langle E^2_{\alpha,\infty}\rangle }$\; \\
\hline
$e$           & $11.12\;{\rm MeV}$ \\  
$\mu$, $\tau$ & $14.99\;{\rm MeV}$ \\  
$\bar{e}$     & $13.65\;{\rm MeV}$ \\   
$\bar{\mu}$, $\bar{\tau}$ & $15.07\;{\rm MeV}$ 
\end{tabular}
\caption{The luminosities, mean energies, and rms energies used for the cooling phase calculation.} \label{tab:coolingphase}
\end{table}

\begin{figure}[t]
\includegraphics[clip,angle=270,width=\linewidth]{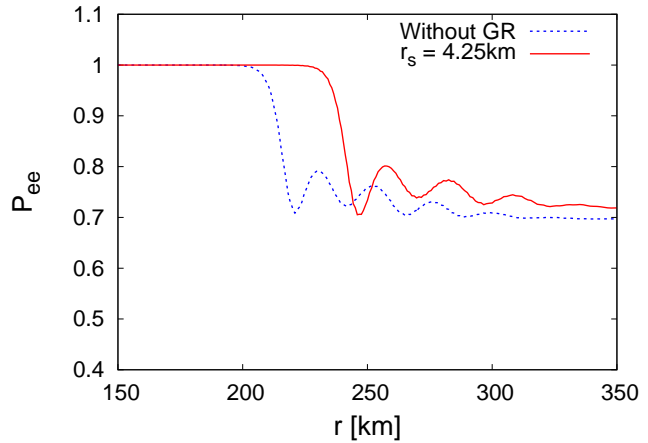}
\caption{The survival probability of electron neutrinos as a function of distance using neutrino spectra and a density profile taken from the cooling phase of a simulation of a $10.8\;{\rm M_{\odot}}$ progenitor by Fischer \emph{et al.} \cite{2010A&A...517A..80F}. The electron flavor survival probability is averaged over all angular bins and energy bins. The red solid line and blue dotted line are the results with and without GR effects respectively.}
\label{fig:cooling_phase}
\end{figure}

The result of this calculation is shown in Fig. \ref{fig:cooling_phase} where we plot the electron neutrino flavor survival probability averaged over all angular bins and energy bins (using the emitted neutrino spectrum as the weighting function) as a function of distance. At this epoch self-interaction effects occur much closer to the proto-neutron star and the effect of GR is more important. The net result of adding GR is to delay the onset of bipolar oscillations by around $25\;{\rm km}$ and once more we find the probability at large radii are almost identical to that without GR. But while this shift in the onset of bipolar oscillations may seem small, we note the neutrino flavor evolution in the region from $50\;{\rm km} \lesssim r \lesssim 500\;{\rm km}$ was found to be crucial for determining the nucleosynthesis yields in the calculations by Duan \emph{et al.} \cite{2011JPhG...38c5201D} and Wu \emph{et al.} \cite{2015PhRvD..91f5016W} so even a relatively small delay of flavor transformation caused by GR might have a consequence.


\section{The GR Neutrino Halo}
\label{sec:halo}

\begin{figure}[t]
\includegraphics[clip,angle=0,width=\linewidth]{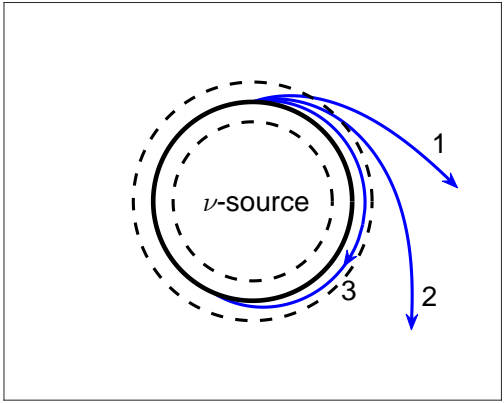}
\caption{Typical neutrino trajectories near a ultra-compact source. The inner dashed lines and the outer dashed lines represent the Schwarzschild radius and the photon sphere respectively. The three trajectories correspond to three different emission angles.}
\label{fig:GR_neutrino_halo}
\end{figure}

So far we have considered only cases where all neutrinos propagate to $r\rightarrow \infty$. However if the compactness of the source becomes too large the neutrinosphere becomes smaller than the ``photon sphere'', whose radius is $3r_s/2$. When this occurs there will be a critical emission angle for neutrinos beyond which they cannot escape to infinity. Following the argument in Hartle \cite{2003gieg.book.....H}, one can obtain a condition for the neutrinos to escape to infinity to be 
\begin{equation}\label{critical_angle}
\frac{2}{{3\sqrt 3 }}\frac{{{R_\nu }}}{{{r_s}}}\frac{1}{{\sqrt {1 - {r_s}/{R_\nu }} }}\sin{\theta_R} < 1.
\end{equation}
We show three example neutrino trajectories for the case where $R_{\nu}/r_s < 3/2 $ in Fig. \ref{fig:GR_neutrino_halo}. Trajectories 1 and 2 are open and a neutrino emitted along these trajectories will propagate to infinity: the trajectories of neutrinos emitted at sufficiently large angles - such as trajectory 3 - will turn around and return to the proto-neutron star. Note that the farthest place where a neutrino can turn around is the photon sphere. The consequence of such trajectories are included in simulations which include GR. In principle there is a substantial change to the flavor evolution calculations when neutrinos start to follow trajectories such as the trajectory 3 in Fig. \ref{fig:GR_neutrino_halo} because they lead to the formation of a neutrino `halo' around the proto-neutron star, similar to the neutrino halos produced by scattering on matter \cite{2012PhRvL.108z1104C,2013PhRvD..87h5037C}. 

\begin{figure}[b]
\includegraphics[clip,angle=0,width=\linewidth]{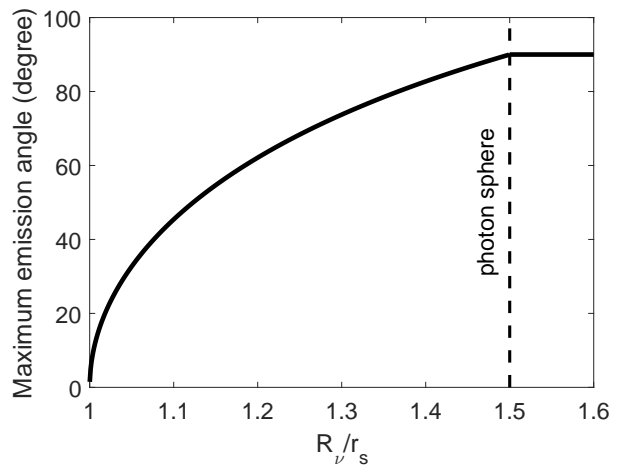}
\caption{The maximum emission angle of neutrinos that can escape the source, for different values of $R_{\nu}/r_s$. The vertical dashed line indicates the position of the photon sphere.}
\label{fig:critical_angle}
\end{figure}

From Eq. (\ref{critical_angle}) we can evaluate the critical angle as a function of $R_{\nu}/r_s$. The relation between the critical angle as a function of $R_{\nu}/r_s$ is shown in Fig. \ref{fig:critical_angle}. If $R_{\nu}/r_s>3/2$, clearly neutrinos with all emission angles can escape and no neutrino halo is formed. We define a critical compactness  $\xi_{\nu\star}$ to be the case where $R_{\nu}/r_s = 3 /2$ and find it equal to $\xi_{\nu\star} = 2.26$ - the value discussed earlier. The compactness of the sources we have considered for our previous numerical calculations did not approach this value because the mass of the proto-neutrons star is not sufficiently large and the neutrinospheres lay beyond the photon sphere. To reach the critical compactness for formation of the halo we require a more massive proto-neutron star with a smaller neutrinosphere.
Whether a proto-neutron star surpasses the critical compactness while the proto-neutron star is still cooling via neutrino emission will depend upon the Equation of State of dense matter and the neutrino opacity \cite{2016PhR...621..127L,2016arXiv161110226H}. 
Note that from causality, the radius of a neutron star is required to be greater than $R_{NS} \gtrsim 2.823 M$ \cite{2016PhR...621..127L} which, if we set $R_{\nu} = R_{NS}$, corresponds to a compactness of $\xi_{\nu} = 2.4$, which is beyond the critical value $\xi_{\nu\star}$. A halo will certainly form immediately preceding the collapse of a proto-neutron star to a black hole. 

The formation of a neutrino halo has consequences for the cooling of the proto-neutron star as well as the flavor transformation due to neutrino self-interaction. One can find a presentation of the changes that occur to the emitted neutrino spectra as the mass of the proto-neutron star approaches its maximum mass in Liebend{\"o}rfer \emph{et al.} \cite{2004ApJS..150..263L}. In their simulations, as the maximum mass is approached (but before the black holes forms) the luminosity of the $\mu$ and $\tau$ flavors increases due to contraction of the proto-neutron star while the luminosities of electron neutrino and electron antineutrinos drop. The mean energies of all flavors increases. 

When a halo forms, in principle, one would have to completely change how the flavor calculations are undertaken in the halo region - the zone between the neutrinosphere and the photon sphere. In such cases the flavor evolution up to the photon sphere cannot be treated as an initial value problem - as we have done in this paper - because the flavor evolution up to the photon sphere of outward moving neutrinos is affected by neutrinos that were also emitted in an outward direction but which turned around and are now moving inwards. Thus in the halo region a paradigm beyond the bulb model would be needed to correctly deal with the flavor evolution. Prevailing understanding from the extant literature would indicate that in the case of three active flavors of neutrino emitted spherically symmetrically, one should not expect flavor transformation within the halo: if this is true then the only effect of the formation of a halo would be to alter the luminosity and angular distribution of the neutrinos beyond the photon sphere (which now becomes the effective neutrinosphere). But in other circumstances - such as calculations that include sterile neutrinos \cite{1999PhRvC..59.2873M,2006PhRvD..73i3007B,2012JCAP...01..013T,2014PhRvD..90j3007W,2014PhRvD..90c3013E} or calculations with non-standard neutrino interactions \cite{2007PhRvD..76e3001E,2008PhRvD..78k3004B,2010PhRvD..81f3003E,2012PhRvD..85k3007S} - flavor transformation can occur much closer to the neutrinosphere in which case the formation of a halo may have greater consequences. 


\section{Summary and Conclusions}
\label{sec:conclusions}

In this paper we have considered the effects of General Relativity upon neutrino flavor transformation in a core-collapse supernova. We adopted a Schwarzschild metric to describe the spacetime and included three GR effects - trajectory bending, time dilation, and energy redshift. Of the three, time dilation is the major effect close to the proto-neutron star, whose role is replaced by trajectory bending at larger radii. The size of the GR effects were found to scale with a single parameter which is the compactness of the source: the relative ratio of the Schwarzschild radius to the neutrinosphere radius. For large compactness with $R_{\nu}$ close to the radius of the photon sphere, the neutrino self-interaction Hamiltonian can be up to approximately three times larger than without GR. We calculated the flavor evolution in two representative cases to determine whether the GR effects led to significant differences compared to calculations without GR. These cases were a density profile and neutrino spectra typical of the accretion phase, and a density profile and neutrino spectra typical of the cooling phase. In both cases we found the effect of GR was to delay the onset of flavor transformation but for the accretion phase the flavor transformation occurred due to decoherence at large radii where the change would have little consequence. In contrast, the change to the onset of bipolar oscillations during the cooling phase may be more important because it is much closer to the proto-neutron star and may impact the nucleosynthesis in the neutrino driven wind. Finally, we showed that GR effects can produce a halo of neutrinos surrounding the proto-neutron star for very compact neutrino sources. If a halo forms then, in principle, one would have to treat flavor transformation in the halo region using a different technique than the usual approach of treating it as an initial-value problem.

\begin{acknowledgments}
This research was supported at NC State University by DOE award DE-FG02-10ER41577.
\end{acknowledgments}


\appendix*
\section{The GR corrected expression for the neutrino self-interaction}
\label{sec:appendix}
In order to get the correct expression for $dn_{\alpha}(r,q,\theta)$, we start from the conservation of neutrino flow through an enclosing spherical surface after taking time dilation into account but ignoring flavor transformation. This allows us to write
 \begin{eqnarray}\label{eqn:F0}
\nonumber {r^2}{\mkern 1mu} \sqrt {B\left( r \right)} {\mkern 1mu} {F_\alpha }\left( {r,q} \right)dq &=& R_\nu ^2{\mkern 1mu} \sqrt {B\left( {{R_\nu }} \right)} {\mkern 1mu} {F_\alpha }\left( {{R_\nu },{q_0}} \right)dq_0, 
\\
\end{eqnarray}
where $F_{\alpha}(r,q)$ is the flux of neutrinos with energy $q$ at $r$ per unit energy that were emitted with energy $q_0$ at the neutrinosphere. 
Integrated over all momenta, both sides of this equation must evaluate to 
$\frac{1}{4\pi}L_{\alpha,\infty} / \left\langle E_{\alpha,\infty}\right\rangle$ where 
$L_{\alpha,\infty}$ is the luminosity of flavor $\alpha$ at infinity assuming no oscillations, and similarly $\left\langle E_{\alpha,\infty}\right\rangle$ is the mean energy at infinity again assuming no oscillations. At the neutrinosphere $R_{\nu}$ we have
\begin{equation}\label{eqn:F1}
F_{\alpha}\left(R_{\nu},q_0 \right) = \int_0^1 2\pi j_{\alpha}\left(q_0,\theta'_{R} \right)\,\cos\theta'_{R}\, d\cos\theta'_{R},
\end{equation}
where $j_{\alpha}\left( q_0,\theta_{R} \right)$ is the emitted intensity of flavor $\alpha$ with energy $q_0$ at angle $\theta_{R}$ with respect to the radial direction. At radial coordinate $r$ the flux is 
\begin{equation}\label{eqn:F2}
F_{\alpha}\left(r,q\right) = \int_0^{\theta_{\max}} \cos \theta'\,dn_{\alpha}\left( r, q,\theta' \right),
\end{equation}
where $\theta_{\max}$ is the angle with respect to the radial direction of neutrinos that were emitted at the neutrinosphere with angle $\theta_R = \pi/2$. 
Combining Eq. (\ref{eqn:F0}),(\ref{eqn:F1}) and (\ref{eqn:F2}) we obtain the result that
\begin{widetext}
\begin{equation}
dn_{\alpha}\left( r,q,\theta \right) = \frac{2\pi R_\nu^2}{r^2}\, \sqrt{\frac{B\left(R_{\nu}\right)}{B\left(r\right)}}\,{j_\alpha }\left(q_0,\theta_{R}\right)\,\left(\frac{\cos\theta_R}{\cos\theta}\right)\left(\frac{dq_0}{dq}\right)\,d\cos\theta_{R},
\end{equation}
In the case of half-isotropic emission the intensity $j_{\alpha}$ is independent of $\theta_R$ and can be written as 
\begin{equation}
j_{\alpha}\left(q_0\right) = \frac{1}{4\,\pi^2\,R_\nu^2\sqrt {B\left(R_{\nu}\right)}}\,\left[\frac{L_{\alpha,\infty} }{\left\langle E_{\alpha,\infty}\right\rangle}\right]\,f_{\alpha}\left( q_0\right),
\end{equation}
where $f_{\alpha}\left( q_0 \right)$ is the normalized spectral distribution for flavor $\alpha$ at $R_{\nu}$. The final expression for $dn_{\alpha }\left( {r,q,\theta} \right)$ is thus
\begin{equation}
dn_{\alpha}\left( r,q,\theta \right) = \frac{1}{2\,\pi\,r^2\sqrt{B\left(r\right)}}\,\left[ \frac{L_{\alpha,\infty}}{\left\langle E_{\alpha,\infty}\right\rangle} \right]\,f_{\alpha}\left( q_0 \right)\,\left(\frac{\cos\theta_R}{\cos\theta}\right)\left(\frac{dq_0}{dq}\right)\,d\cos\theta_{R}.
\end{equation}
\end{widetext}

\bibliographystyle{apsrev4-1}
\bibliography{GR}

\end{document}